\documentclass{article}
\usepackage[dvips]{epsfig}
\usepackage{graphicx}
\usepackage{amsmath}

\input{tcilatex}

\begin{document}

\begin{center}
\bigskip \makeatletter

\makeatother
\makeatletter

{\LARGE Nonlinear QED and Physical Lorentz Invariance}{\LARGE \ \vspace{15pt}%
\\[0pt]
}

\textbf{A.T. Azatov}$^{1,2}$\textbf{\ and J.L.~Chkareuli}$^{1}$\\[5pt]
$^{1}$\textit{Institute of Physics, Georgian Academy of Sciences, 0177
Tbilisi, Georgia} \vspace{0pt}\\[0pt]

$^{2}$\vspace{0pt}\textit{Department of Physics, University of Maryland,
College Park, MD 20742, USA}\vspace{0pt}\\[0pt]

\bigskip

\bigskip

\bigskip

\bigskip

\bigskip \bigskip \vspace{55pt} \texttt{Abstract}
\end{center}

\bigskip

The spontaneous breakdown of 4-dimensional Lorentz invariance in the
framework of QED\ with the nonlinear vector potential constraint \ $A_{\mu
}^{2}=M^{2}$\ (where $M$ is a proposed scale of the Lorentz violation) is
shown to manifest itself only as some noncovariant gauge choice in the
otherwise gauge invariant (and Lorentz invariant) electromagnetic theory.
All the contributions to the photon-photon, photon-fermion and
fermion-fermion interactions violating the physical Lorentz invariance
happen to be exactly cancelled with each other in the manner observed by
Nambu a long ago for the simplest tree-order diagrams - the fact which we
extend now to the one-loop approximation and for both\ the time-like ($%
M^{2}>0$) and space-like ($M^{2}<0$) Lorentz viola tion. The way
how to reach the physical breaking of the Lorentz invariance in
the pure QED case (and beyond) treated in the flat Minkowskian
space-time is also discussed in
some detail. 
\thispagestyle{empty} \newpage\

\section{Introduction}

Spontaneous violation of Lorentz invariance has attracted considerable
attention in the last years as an interesting phenomenological possibility
appearing in the framework of various quantum field and string theories \cite
{alan,jakiw,glashow,mota}. For spontaneous Lorentz invariance violation
(LIV), the situation is in some sense similar to the internal symmetry
breaking with the corresponding massless Nambu-Goldstone modes appeared. For
the LIV such modes are believed to be photons\ or even non-Abelian gauge
fields \cite{bjorken}, if the starting symmetry in the Lagrangian is
properly chosen.

The handy theoretical laboratory for these considerations happens to be some
simple class of the Lagrangian models for the starting massive vector field $%
A_{\mu }$ where, in one way or another, the nonlinear dynamical constraint
of type
\begin{equation}
\text{\ }A_{\mu }^{2}=M^{2}\text{\ }  \label{constr}
\end{equation}
($M$ is a proposed scale of the LIV) is appeared. This constraint means in
essence the vector field $A_{\mu }$ develops the vacuum expectation value
(VEV) and Lorentz symmetry $SO(1,3)$ formally breaks down to $SO(3)$ or $%
SO(1,2)$ depending on the sign of the $M^{2}$. Such models, which is often
called \ `bumblebee models' in the literature\cite{alan}, were introduced by
Dirac\cite{dirac} in the fifties (though in a different context) and then
from the LIV point of view was studied by Nambu \cite{nambu} (see also\cite
{venturi}) independently of the dynamical mechanism which causes the
spontaneous Lorentz violation. For this purpose he applied the technique of
nonlinear symmetry realizations which appeared successful in handling the
spontaneous breakdown of chiral symmetry, particularly, as it appears in the
nonlinear $\sigma $ model\cite{GL}. It was shown, while only in the tree
approximation and for the time-like LIV ($M^{2}>0$), that the non-linear
constraint (\ref{constr}) implemented into standard QED Lagrangian
containing the charged\ ($e$) fermion $\psi (x)$%
\begin{equation}
\mathcal{L}_{QED}=-\frac{1}{4}F_{\mu \nu }F^{\mu \nu }+\overline{\psi }%
(i\gamma \partial +m)\psi -eA_{\mu }\overline{\psi }\gamma ^{\mu }\psi \text{
\ \ }  \label{lagr1}
\end{equation}
as some supplementary condition appears in fact as a possible gauge choice
which amounts to a temporal gauge for the superlarge (as it is intuitively
expected) LIV scale $M$. At the same time, the $S$-matrix remains unaltered
under such a gauge convention. This particular gauge allows one \ to
interpret QED in terms of the spontaneous LIV with the VEV of vector field
of the type $<A_{\mu }>_{0}$ $=(M,0,0,0)$. The LIV, however, is proved to be
superficial as it affects only the gauge of vector potential $A_{\mu }$ at
least in the tree approximation \cite{nambu}.

In this connection it is a matter of great importance to know whether the
Nambu's observation remains when quantum corrections are included into
Lagrangian (\ref{lagr1}). One might think that the tree LIV diagrams are
actually cancelled since this level corresponds in fact to the classical
theory where the constraint (\ref{constr}) manifests itself as a pure gauge.
However, including into play the loop diagrams, which means that one comes
to the quantum theory where the vector field canonical commutators
introduced (being the non-trivial constraints by themselves), could not
allow to consider further the constraint $A_{\mu }^{2}=M^{2}$ as a gauge
choice and, as result, the physically observable LIV effects might appear.

We are focused here on the lowest order LIV processes in QED with the
nonlinear dynamical constraint (\ref{constr}) for both of cases of the
time-like ($M^{2}>0$) and space-like ($M^{2}>0$) LIV. We explicitly show
that for tree approximation all the LIV contributions are exactly cancelled
with each other just in a manner which was observed by Nambu a long ago. We
then extend our consideration to the calculation of the one-loop LIV\
contributions to the photon-photon, photon-fermion and fermion-fermion
scattering. All these contributions are shown to be mutually cancelled in
the framework of the particular dimensional regularization scheme taken (in
the way as this scheme is usually applied to QED in noncovariant gauges\cite
{leib}). This means that the constraint $A_{\mu }^{2}=M^{2}$ having been
treated as the nonlinear gauge choice at a tree (classical) level remains as
a gauge condition when quantum effects are taken into account as well. So,
in accordance with Nambu's original conjecture one can conclude that\ the
physical Lorentz invariance is left intact at least in the one-loop
approximation provided we consider the standard QED Lagrangian (\ref{lagr1}
(with its gauge invariant $F_{\mu \nu }F^{\mu \nu }$ kinetic term and
minimal photon-fermion coupling) taken in the flat Minkowskian space-time.

The paper is organized in the following way. We consider first the
non-linear QED Lagrangian (Sec.2) appeared once the dynamical constraint (%
\ref{constr}) is explicitly implemented into Lagrangian (\ref{lagr1}), and
derive the general Feynman rules for the basic photon-photon and
photon-fermion interactions depending no on the particular case of the
time-like or space-like LIV. The model appears in essence two-parametric
containing the electric charge $e$ and inverse LIV scale $1/M$ as the
perturbation parameters so that the LIV interactions are always proportional
some powers of them. Then in Sec.3 the LIV tree processes are discussed and,
as a typical example, the Lorentz violating Compton effect in the lowest $%
e/M $ order is considered in detail. In addition to Nambu's conclusion, we
have shown that the total cancellation of the physical LIV tree effects has
place in both of cases $M^{2}>0$ and $M^{2}<0$. In Sec.4 we present the
detailed calculation of the one-loop contributions to the fermion-fermion
scattering in the $e^{3}/M$ order \ and also briefly discuss the other
leading one-loop contributions to the photon-photon, photon-fermion and
fermion-fermion scattering up to the next LIV order $1/M^{2}$. All these
effects appear in fact vanishing. Actually, their matrix elements, when they
do not vanish by themselves, amount to the differences between pairs of the
similar integrals whose integration variables are shifted relative to each
other by some constants (being in general arbitrary functions of the
external four-momenta of the particles involved) that in the framework of
the dimensional regularization leads to their total cancellation. And,
finally, we give our conclusions in Sec.5. Among them the way how to reach
the physical breaking of Lorentz invariance in the flat Minkowskian
space-time is also discussed in some detail.

\section{\protect\bigskip The Lagrangian and Feynman rules}

\subsection{\textbf{\ The Lagrangian}}

We consider simultaneously both of the above-mentioned LIV cases, time-like
or space-like, introducing some unit vector $n_{\mu }$ ($n_{\mu }^{2}\equiv
n^{2}=\pm 1$ depending on the sign of $\ M^{2}$, respectively) so as to have
the following general parametrization for the vector potential $A_{\mu }$ in
the Lagrangian (\ref{lagr1}) of the type
\begin{equation}
\text{\ \ }A_{\mu }=a_{\mu }+\frac{n_{\mu }}{n^{2}}(n\cdot A)  \label{par}
\end{equation}
where the $a_{\mu }$ is pure Goldstonic mode
\begin{equation}
\text{\ }n\cdot a=0\text{\ }  \label{sup}
\end{equation}
while the Higgs mode (or the $A_{\mu }$ component in the vacuum direction)
is given by the scalar product $n\cdot A$. Substituting this parametrization
into the vector field constraint (\ref{constr}) one comes to the equation
for $n\cdot A$ (taking, for simplicity, the positive sign for the square
root only)
\begin{equation}
\text{\ }n\cdot A\text{\ }=\left[ n^{2}(M^{2}-a_{\nu }^{2})\right] ^{\frac{1%
}{2}}  \label{constr1}
\end{equation}
which for the particular time-like ($M^{2}>0,$ $n^{2}=1$) and space-like ($%
M^{2}<0,$ $n^{2}=-1$) VEV cases takes the simpler forms
\begin{equation}
\text{\ }A_{0}\text{\ }=\left[ M^{2}+a_{i}^{2})\right] ^{\frac{1}{2}},\text{
\ \ \ }n_{\mu }=(1,0,0,0),\text{ \ \ \ }a_{\mu }=(0,A_{i}),\text{\ \ }%
(i=1,2,3)  \label{constr2}
\end{equation}
and
\begin{equation}
\text{\ }A_{3}\text{\ }=\left[ \left| M^{2}\right| +a_{\beta }^{2}\right] ^{%
\frac{1}{2}},\text{ }\ \ \text{\ }n_{\mu }=(0,0,0,1),\text{ \ \ \ }a_{\mu
}=(A_{\beta },0),\text{\ \ }(\beta =0,1,2)  \label{constr3}
\end{equation}
respectively (for the space-like case the vacuum direction was chosen along
the third axis). For the high LIV scale $M$, as is expected, the equation
for $n\cdot A$ (\ref{constr1}) can be then expanded in powers of $\frac{%
a_{\nu }^{2}}{M^{2}}$
\begin{equation}
\text{\ }n\cdot A\text{\ }=M-\frac{n^{2}a_{\nu }^{2}}{2M}+O(1/M^{2})
\label{constr4}
\end{equation}
where $M$ is defined always positive, while $n^{^{2}}\equiv n_{\mu
}^{2}=n_{0}^{2}-n_{i}^{2}$ and $a_{\nu }^{2}$ $=a_{0}^{2}-a_{i}^{2}$ are
determined according their non-zero components given in Eqs. (\ref{constr2})
and (\ref{constr3}) for particular cases.

We proceed further putting that new parametrization (\ref{par}) into our
basic Lagrangian (\ref{lagr1}), using then the above expansion for the Higgs
mode $n\cdot A$\ (\ref{constr4}) and making the appropriate redefinition of
fermion field $\psi $ according to
\begin{equation}
\psi \rightarrow e^{ieM(n\cdot x)}\psi
\end{equation}
so that the mass-type term $eM\overline{\psi }(\gamma \cdot n)\psi $
appearing from the expansion of the fermion current interaction in the
Lagrangian (\ref{lagr1}) will be exactly cancelled by an analogous term
stemming now from the fermion kinetic term. So, we eventually arrive at the
Lagrangian for the $a_{\mu }$ field (denoting its strength tensor by $f_{\mu
\nu }=\partial _{\mu }a_{\nu }-\partial _{\nu }a_{\mu }$)

\begin{eqnarray}
\mathcal{L}_{NL} &=&-\frac{1}{4}f_{\mu \nu }f^{\mu \nu }+\overline{\psi }%
(i\gamma \partial +m)\psi -ea_{\mu }\overline{\psi }\gamma ^{\mu }\psi +
\notag \\
&&-\frac{1}{4M}f_{\mu \nu }\left[ \left( n^{\mu }\partial ^{\nu }-n^{\nu
}\partial ^{\mu }\right) a_{\rho }^{2}n_{\sigma }^{2}+\cdot \cdot \cdot %
\right] -\frac{1}{16M^{2}}\left[ \left( n^{\mu }\partial ^{\nu }-n^{\nu
}\partial ^{\mu }\right) a_{\rho }^{2}+\cdot \cdot \cdot \right] ^{2}+
\notag \\
&&+e\frac{a_{\rho }^{2}n_{\sigma }^{2}}{2M}(1+\cdot \cdot \cdot )\overline{%
\psi }(\gamma \cdot n)\overline{\psi }\text{ \ \ }\   \label{lagr2}
\end{eqnarray}
where we collected the linear and nonlinear (in the $a_{\mu }$ fields) terms
separately leaving only terms corresponding to the expansion in the Higgs
mode $n\cdot A$, as is taken in Eq.(\ref{constr4}), and also retained the
former notation for fermion $\psi $. We take the Greek letters for the
Lorentz indices ($\mu ,\nu ,\rho ,\sigma =0,1,2,3$) and the metric is \ $%
g_{\mu \nu }=(1,-1,-1,-1)$, while everywhere when appears $(n^{2})^{2}$ (and
higher powers of $n^{2}$) we replace it by $1$. For the photon-electron and
photon-photon interactions it follows then in the lowest approximation
\begin{equation}
\mathcal{L}_{NL}^{int}=-ea_{\mu }\overline{\psi }\gamma ^{\mu }\psi +e\frac{%
n^{2}a_{\rho }^{2}}{2M}\bar{\psi}(\gamma \cdot n)\psi -\text{ }\frac{n^{2}}{M%
}(\partial _{\mu }a_{\nu }n^{\mu }a_{\rho }\partial ^{\nu }a^{\rho })\text{ }%
-\frac{1}{16M^{2}}\left[ \left( n^{\mu }\partial ^{\nu }-n^{\nu }\partial
^{\mu }\right) a_{\rho }^{2}\right] ^{2}\text{\ }  \label{lagr3}
\end{equation}

The Lagrangian (\ref{lagr2}) together with the gauge fixing condition (\ref
{sup}) completes the nonlinear $\sigma $ model type construction for quantum
electrodynamics. We call this the nonlinear QED. The model contains the
massless vector Goldstone boson modes and keeps the massive Higgs mode
frozen, and in the limit $M\rightarrow \infty $ the model (given just by the
first line in the Lagrangian $\mathcal{L}_{NL}$ (\ref{lagr2})) is
indistinguishable from conventional QED taken in the temporal or axial gauge
(\ref{sup}). So, for this part of the Lagrangian $\mathcal{L}_{NL}$ the
spontaneous LIV only means the noncovariant gauge choice (\ref{sup}) in
otherwise the gauge invariant (and Lorentz invariant) theory. However, we
will show in the next section that also all other terms in the $\mathcal{L}%
_{NL}$ (\ref{lagr2}), though being by themselves the Lorentz and $C(CPT)$
violating ones, cause no the physical LIV effects at least in the one-loop
approximation.

\subsection{\textbf{\ The Feynman rules}}

They for the interaction Lagrangian $\mathcal{L}_{NL}^{int}$ (\ref{lagr3})
include:

i/ An ordinary QED photon-electron vertex is $\ $%
\begin{equation}
-ie\gamma _{\mu }
\end{equation}

ii/ The contact 2-photon-electron vertex is given by $\ \ $%
\begin{equation}
i\frac{eg_{\mu \nu }n^{2}}{M}(\gamma \cdot n)
\end{equation}

iii/ The 3-photon vertex (with photon 4-momenta $k_{1},$ $k_{2}$ and $k_{3}$%
) is appeared as
\begin{equation}
-\frac{in^{2}}{M}\left[ (k_{1}\cdot n)k_{1,\alpha }g_{\beta \gamma
}+(k_{2}\cdot n)k_{2,\beta }g_{\alpha \gamma }+(k_{3}\cdot n)k_{3,\gamma
}g_{\alpha \beta }\right]
\end{equation}
where the second index in the each momentum $k_{1},$ $k_{2}$ and $k_{3}$
denotes its Lorentz component;

iv/ The 4-photon vertex (with photon 4-momenta $k_{1},$ $k_{2},$ $k_{3}$ and
$k_{4}$) is
\begin{eqnarray}
&&\frac{-i}{M^{2}}\{[(n^{2}(k_{1}+k_{2})^{2}-(n(k_{1}+k_{2}))^{2}]g_{\alpha
\beta }g_{\mu \nu }+[n^{2}(k_{1}+k_{3})^{2}-(n(k_{1}+k_{3}))^{2}]g_{\alpha
\mu }g_{\beta \nu }+  \notag \\
&&+[n^{2}(k_{1}+k_{4})^{2}-(n(k_{1}+k_{4}))^{2}]g_{\alpha \nu }g_{\beta \mu
}\}
\end{eqnarray}

v/ The photon propagator is in general (for $n^{\mu }a_{\mu }=0$)
\begin{equation}
D_{\mu \nu }(k)=\frac{-i}{k^{2}+i\epsilon }\left( g_{\mu \nu }-\frac{n_{\mu
}k_{\nu }+k_{\mu }n_{\nu }}{n\cdot k}+\frac{n^{2}k_{\mu }k_{\nu }}{(n\cdot
k)^{2}}\right)  \label{prop}
\end{equation}
(where $k^{2}$ stands for the photon 4-momentum squared) being automatically
satisfied the orthogonality condition $n^{\mu }D_{\mu \nu }(k)=0$ and
on-shell transversality $k_{\mu }D_{\mu \nu }(k)=0$ $\ $ ($k^{2}=0$). The
latter means that the free photon with the polarization vector $\varepsilon
^{\mu }(k,k^{2}=0)$ is always appeared transverse $k_{\mu }\varepsilon ^{\mu
}(k)=0$.

vi/ The electron propagator (standard) is

\begin{equation}
S(p)=\frac{i}{\gamma \cdot p-m}
\end{equation}

\section{The tree LIV contribution: photon-fermion scattering}

\begin{figure}[h]
\begin{center}
\includegraphics[scale=0.3,angle=90]{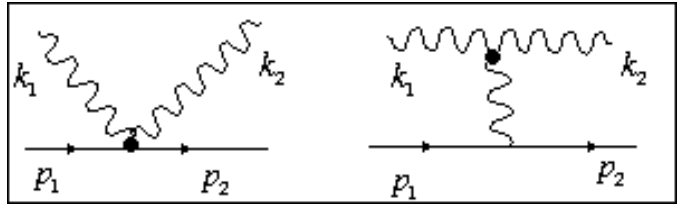}
\end{center}
\caption{The LIV contribution to the Compton effect in the tree
approximation ( in the order of O($\frac{e}{M}$)) }
\label{comptem}
\end{figure}

{We start with a calculation of the tree LIV contributions to the
photon-fermion scattering. We show now that }such contributions to the
standard Compton effect taken in the lowest $e/M$ \ order are exactly
cancelled for any choice of the constant vector $n^{\mu }$ (or for the
time-like or the space-like LIV). These contributions are given by two
diagrams (see Fig.\ref{comptem}).

\subsection{The LIV matrix element}

This matrix element $i\mathcal{M}$ corresponding to these two diagrams is
given by
\begin{equation}
i\mathcal{M}=[\bar{u}(p_{2})(p_{2})O_{\mu \nu }u(p_{1})]\cdot \varepsilon
^{\mu }(k_{1})\varepsilon ^{\nu }(k_{2})
\end{equation}
where the ingoing and outgoing electron spinors $u(p_{1})$ and $u(p_{2})$
(with momenta $p_{1}$ and $p_{2}$) and\ photon polarization vectors $%
\varepsilon _{\mu }(k_{1})$ and $\varepsilon _{\nu }(k_{2})$\ (with momenta $%
k_{1}$and\ $k_{2}$) are explicitly indicated. The $O^{\mu \nu }$ consists of
\ sum of both of diagrams and is in fact
\begin{eqnarray}
O_{\mu \nu } &=&\ i\frac{eg_{\mu \nu }n^{2}}{M}(\gamma n)-\frac{in^{2}}{M}[%
(k\cdot n)k_{\beta }g_{\mu \nu }+(k_{2}n)k_{2,\nu }g_{\mu \beta
}+(k_{1}n)k_{1,\mu }g_{\beta \nu })]\cdot  \notag \\
&&\cdot (-ie\gamma _{\alpha })\frac{(-i)}{k^{2}}\left( g^{\alpha \beta }-%
\frac{n^{\alpha }k^{\beta }+k^{\alpha }n^{\beta }}{n\cdot k}+\frac{%
n^{2}k^{\alpha }k^{\beta }}{(n\cdot k)^{2}}\right)
\end{eqnarray}
where $k$ is the transferred 4-momentum $k=p_{2}-p_{1}=k_{1}-k_{2}$.

\subsection{Cancellation of the tree LIV contributions}

Since the ingoing and outgoing photons appear transverse ( $k_{1,2}\cdot
\varepsilon (k_{1,2})=0$) there are only left the terms
\begin{equation}
i\mathcal{M}=i\frac{en^{2}g_{\mu \nu }}{M}\bar{u}(p_{2})\left[ (\gamma \cdot
n)+\frac{(kn)k_{\beta }\gamma _{\alpha }}{k^{2}}\left( g^{\alpha \beta }-%
\frac{n^{\alpha }k^{\beta }+k^{\alpha }n^{\beta }}{(n\cdot k)}+\frac{%
n^{2}k^{\alpha }k^{\beta }}{(n\cdot k)^{2}}\right) \right] u(p_{1})\cdot
\varepsilon ^{\mu }(k_{1})\varepsilon ^{\nu }(k_{2})
\end{equation}
in the matrix element $i\mathcal{M}$. So, after the evident simplification
in the square bracket
\begin{equation}
(\gamma \cdot n)+\frac{(k\cdot n)k^{\alpha }\gamma _{\alpha }}{k^{2}}+\frac{%
n^{2}k^{\alpha }\gamma _{\alpha }}{(n\cdot k)}-(\gamma n)-\frac{%
(kn)k^{\alpha }\gamma _{\alpha }}{k^{2}}=\frac{n^{2}\hat{k}}{(n\cdot k)}
\end{equation}
one is finally led to the matrix element ($k=p_{2}-p_{1}$)
\begin{equation}
i\mathcal{M}=i\frac{en^{2}}{M}\left[ \bar{\psi}(p_{2})\frac{\hat{p}_{2}-\hat{%
p}_{1}}{n\cdot (p_{2}-p_{1})}\psi (p_{1})\right] \cdot \lbrack \varepsilon
(k_{1})\cdot \varepsilon (k_{2})]
\end{equation}
which unavoidably amounts to zero due to the fermionic current conservation
\begin{equation}
\bar{u}(p_{2})(\hat{p}_{2}-\hat{p}_{1}){u(p_{1})=0}
\end{equation}

\subsection{The other tree LIV\ processes}

We have also considered the other processes in the tree approximation, such
as the pure photon-photon scattering (going through the pole 3-photon and
contact 4-photon diagrams), electron-electron scattering with emission of
extra photon ($e+e\rightarrow e+e+\gamma $) etc. taken in the lowest order,
and everywhere the LIV contributions are completely cancelled. Moreover, in
addition to the Nambu's conclusion we found that such a cancellation has
place for both of signs of $M^{2}$, as one can readily see from the
above-considered Compton scattering case (i.e. the cancellation occurs for
any choice of the vector $n_{\mu }$). It seems very likely that such
tendency remains in the higher-order tree diagrams as well, since there
works the special mechanism of cancellation between the 3-photon diagram and
the corresponding contact diagram (like as we explicitly showed for the
Compton scattering diagrams). Remarkably, the same mechanism of cancellation
happens to also work for the loop diagrams, as we can see in the next
section.

\section{\protect\bigskip The loop LIV contribution: fermion-fermion
scattering}

Consideration of the loop LIV contributions appears much more complicated
since the nonlinear QED (\ref{lagr2}) seems to be (at least formally)
nonrenormalizable theory in which even the one-loop divergences could be
gauge dependent. In this connection the choice of an adequate regularization
scheme in the model is a matter of a crucial importance. As the typical
process including the one-loop LIV corrections we consider in detail the $%
ee^{\prime }$ scattering process in the lowest $e^{3}/M$ order \ and also
briefly discussed the proper one-loop contributions to the photon-photon,
photon-fermion and\ fermion-fermion scattering up to the higher LIV order $%
1/M^{2}$. The $e^{\prime }$ could be any other lepton, say, muon $\mu $ or
taon $\tau $, so that the complications related with the identical fermions
are avoided. We show here the LIV cancellation mechanism, which appeared so
effective in the above for the tree LIV\ diagrams, happens to work for the
loop contributions as well in the framework of the dimensional
regularization scheme taken. In that scheme the possible surface terms
appearing from the (linearly and higher) divergent integrals in the model
automatically vanish thus allowing the LIV cancellation mechanism to work
without serious consequences. At the same time one could apply some other
regularization which would feel such surface terms and, as a result, some
surviving physical LIV effects could appear. We discuss this point for the $%
ee^{\prime }$ scattering process in detail at the end of this section

\subsection{\protect\bigskip Basic diagrams and matrix element}

\begin{figure}[h]
\begin{center}
\includegraphics[scale=0.3]{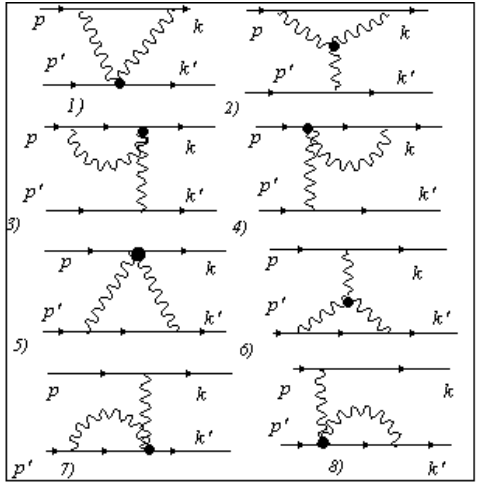}
\end{center}
\caption{The LIV contribution to the $e-e^{\prime }$ scattering in the
one-loop approximation ( in the order of O($\frac{e^{3}}{M}$)) }
\label{emuem3}
\end{figure}

The basic LIV\ diagrams for the $ee^{\prime }$ scattering stem from the
interaction Lagrangian (\ref{lagr2}) properly extended to include the lepton
$e^{\prime }$ as well. There are in fact eight possible diagrams in the
lowest order $e^{3}/M$, as are given in the Fig.\ref{emuem3}.
According to them and Feynman rules formulated in Sec.2.2 one immediately
finds the corresponding matrix element. Actually, we consider four diagrams
(1+2+3+4) (the contribution of the other four diagrams (5+6+7+8) follows
from the simple replacement of 4-momenta $p\Leftrightarrow p^{\prime }$ and $%
k\Leftrightarrow k^{\prime }$ and masses $m\Leftrightarrow m^{\prime }$ in
the matrix element $i\mathcal{M}_{1-4}$). Considering first the diagrams
(1+2) one has
\begin{eqnarray}
i\mathcal{M}_{1+2} &=&\bar{u}(k^{\prime }){\large \{}\ i\frac{eg_{\alpha
\beta }n^{2}}{M}(\gamma \cdot n)+ \\
&&+(-ie\gamma _{\lambda })D^{\lambda \rho }(p-k)(\frac{-in^{2}}{M}){\large [}%
[(p-k)\cdot n](p-k)_{\rho }g_{\alpha \beta }+  \notag \\
&&+[(p-q)\cdot n](p-q)_{\alpha }g_{\beta \rho }+[(k-q)\cdot n](k-q)_{\beta
}g_{\alpha \rho }{\large ]\}}u(p^{\prime })\cdot  \notag \\
&&\cdot {\large \lbrack }\bar{u}(k)D^{\beta \nu }(k-q)(-ie\gamma _{\nu })(i)%
\frac{\hat{q}+m}{q^{2}-m^{2}}D^{\alpha \mu }(p-q)(-ie\gamma _{\mu })u(p)%
{\large ]}  \notag
\end{eqnarray}
where the total energy-momentum conservation $p+p\prime =k+k\prime $ (so
that $D^{\lambda \rho }(p-k)=D^{\lambda \rho }(p\prime -k\prime )$ etc.), as
well as the standard integration with respect to the internal 4-momentum $q$
($i\mathcal{M}_{1+2}\rightarrow \int \frac{d^{4}q}{(2\pi )^{4}}(i\mathcal{M}%
_{1+2})$) are also implied.

\bigskip

\subsection{Cancellation of the loop LIV contributions}

One can see now that the contact $\gamma \cdot n$ term in $i\mathcal{M}%
_{1+2} $ is cancelled with the first term in the square bracket (containing
the 3-photon vertex terms) since the sum of these two terms can be rewritten
as (using the photon propagator form (\ref{prop}) for $D^{\lambda \rho
}(p-k) $)
\begin{eqnarray}
&&i\frac{eg_{\alpha \beta }n^{2}}{M}\bar{u}(k^{\prime })\{(\gamma \cdot
n)+ie\gamma _{\lambda }D^{\lambda \rho }(p-k)[(p-k)\cdot n](p-k)_{\rho
}\}u(p^{\prime }))= \\
&&i\frac{eg_{\alpha \beta }n^{2}}{M}\bar{u}(k^{\prime }){\LARGE \{}(\gamma
\cdot n)+\frac{[(p-k)\cdot n](\hat{p}-\hat{k})}{(p-k)^{2}}+\frac{n^{2}(\hat{p%
}-\hat{k})}{(p-k)\cdot n}-(\gamma \cdot n)-\frac{((p-k)\cdot n)(\hat{p}-\hat{%
k})}{(p-k)^{2}}{\LARGE \}}u(p^{\prime })\text{ }  \notag \\
&=&i\frac{eg_{\alpha \beta }}{M}\bar{u}(k^{\prime })\frac{\hat{p}-\hat{k}}{%
(p-k)\cdot n}u(p^{\prime })  \notag
\end{eqnarray}
which due to the conservation of fermion current (and 4-momentum
conservation $p-k=-(p^{\prime }-k^{\prime })$) certainly comes to zero. This
is just the general mechanism already found in the Compton effect (see
Sec.3): the contact 2-photon diagram contribution is always cancelled with
the part of the pole 3-photon diagram contribution which is free from the
internal integration.

At the same time its other parts still remain. They correspond just to two
survived 3-photon vertex terms in the $i\mathcal{M}_{1+2}$ which are subject
to an integration with respect to the internal 4-momentum $q$. They amount
to (properly rewritten)
\begin{eqnarray}
i\mathcal{M}_{1+2} &=&\frac{ie^{3}n^{2}}{M}{\LARGE [}\bar{u}(k^{\prime
})\gamma _{\lambda }u(p^{\prime }){\LARGE ]}D^{\lambda \rho }(p-k){\LARGE \{}%
[(p-q)\cdot n](p-q)_{\alpha }D^{\alpha \mu }(p-q)g_{\beta \rho }D^{\beta \nu
}(k-q)+  \notag \\
&&+[(k-q)\cdot n](k-q)_{\beta }D^{\beta \nu }(k-q)g_{\alpha \rho }D^{\alpha
\mu }(p-q){\LARGE \}[}\bar{u}(k)\gamma _{\nu }\frac{\hat{q}+m}{q^{2}-m^{2}}%
\gamma _{\mu }u(p){\LARGE ]}
\end{eqnarray}
Using there
\begin{eqnarray}
\lbrack (p-q)\cdot n](p-q)_{\alpha }D^{\alpha \mu }(p-q) &=&(-i){\LARGE [}%
\frac{n^{2}(p-q)^{\mu }}{(p-q)\cdot n}-n^{\mu }{\LARGE ]} \\
\lbrack (k-q)\cdot n](k-q)_{\beta }D^{\beta \nu }(k-q) &=&(-i){\LARGE [}%
\frac{n^{2}(k-q)^{\nu }}{(k-q)\cdot n}-n^{\nu }{\LARGE ]}  \notag
\end{eqnarray}
and then (using Dirac equations $\hat{p}\cdot u(p)=m\cdot u(p)$, $\bar{u}%
(k)\cdot \hat{k}=\bar{u}(k)\cdot m$ \ in the fermionic sector)
\begin{eqnarray}
(p-q)^{\mu }{\LARGE [}\bar{u}(k)\gamma _{\nu }\frac{\hat{q}+m}{q^{2}-m^{2}}%
\gamma _{\mu }u(p){\LARGE ]} &=&-\bar{u}(k)\gamma _{\nu }u(p) \\
(k-q)^{\nu }{\LARGE [}\bar{u}(k)\gamma _{\nu }\frac{\hat{q}+m}{q^{2}-m^{2}}%
\gamma _{\mu }u(p){\LARGE ]} &=&-\bar{u}(k)\gamma _{\mu }u(p)  \notag
\end{eqnarray}
one is finally led to
\begin{eqnarray}
i\mathcal{M}_{1+2} &=&\frac{e^{3}n^{2}}{M}{\large [}\bar{u}(k^{\prime
})\gamma _{\lambda }u(p^{\prime })][\bar{u}(k)\gamma _{\mu }u(p){\large ]}%
D^{\lambda \rho }(p-k)D_{\rho }^{\mu }(k-q)\frac{1}{(q-p)\cdot n}+  \notag \\
&&+\frac{e^{3}n^{2}}{M}{\large [}\bar{u}(k^{\prime })\gamma _{\lambda
}u(p^{\prime }){\large ][}\bar{u}(k)\gamma _{\mu }u(p){\large ]}D^{\lambda
\rho }(p-k)D_{\rho }^{\mu }(p-q)\frac{1}{(q-k)\cdot n}-  \notag \\
&&-\frac{e^{3}n^{2}}{M}{\large [}\bar{u}(k^{\prime })\gamma _{\lambda
}u(p^{\prime }){\large ][}\bar{u}(k)\gamma _{\mu }\frac{\hat{q}+m}{%
q^{2}-m^{2}}(\gamma \cdot n)u(p){\large ]}D^{\lambda \rho }(p-k)D_{\rho
}^{\mu }(k-q)-  \label{1+2} \\
&&-\frac{e^{3}n^{2}}{M}{\large [}\bar{u}(k^{\prime })\gamma _{\lambda
}u(p^{\prime }){\large ][}\bar{u}(k)(\gamma \cdot n)\frac{\hat{q}+m}{%
q^{2}-m^{2}}\gamma _{\mu }u(p){\large ]}D^{\lambda \rho }(p-k)D_{\rho }^{\mu
}(p-q)  \notag
\end{eqnarray}
where $D_{\rho }^{\mu }(k-q)=g_{\alpha \rho }D^{\alpha \mu }(k-q)$ etc.

Let us turn to the diagrams (3) and (4) in the Fig.2. Their calculation goes
faster and one readily has

\begin{eqnarray}
i\mathcal{M}_{3+4} &=&\bar{u}(k^{\prime })(-ie\gamma _{\lambda })D^{\lambda
\rho }(p-k)u(p^{\prime })\cdot  \label{3+4} \\
&&\cdot \bar{u}(k){\large [}(-ie\gamma _{\mu })(i)\frac{\hat{q}+m}{%
q^{2}-m^{2}}D^{\mu \alpha }(k-q)(i\frac{eg_{\alpha \rho }n^{2}}{M})(\gamma
\cdot n)+  \notag \\
&&+(i\frac{eg_{\alpha \rho }n^{2}}{M})(\gamma \cdot n)D^{\mu \alpha }(k-q)(i)%
\frac{\hat{q}+m}{q^{2}-m^{2}}(-ie\gamma _{\mu }){\large ]}u(p)  \notag \\
&=&\frac{e^{3}n^{2}}{M}{\large [}\bar{u}(k^{\prime })\gamma _{\lambda
}u(p^{\prime }){\large ][}\bar{u}(k)\gamma _{\nu }\frac{\hat{q}+m}{%
q^{2}-m^{2}}(\gamma \cdot n)u(p){\large ]}D^{\lambda \rho }(p-k)D_{\rho
}^{\nu }(k-q)+  \notag \\
&&+\frac{e^{3}n^{2}}{M}{\large [}\bar{u}(k^{\prime })\gamma _{\lambda
}u(p^{\prime }){\large ][}\bar{u}(k)(\gamma \cdot n)\frac{\hat{q}+m}{%
q^{2}-m^{2}}\gamma _{\mu }u(p){\large ]}D^{\lambda \rho }(p-k)D_{\rho }^{\mu
}(p-q)  \notag
\end{eqnarray}
One can immediately see that the $i\mathcal{M}_{3+4}$ is completely
cancelled with the last two terms in the $i\mathcal{M}_{1+2}$ (\ref{1+2}).
Strictly speaking these terms in the $i\mathcal{M}_{3+4}$ (\ref{3+4}) can
differ (as being followed from the different diagrams) from the
corresponding terms in the $i\mathcal{M}_{1+2}$ by some arbitrary shifts in
the integration variable $q$. One could take instead, say, $q$ $+a$ and $q$ $%
+b$ in the\ integrals in $i\mathcal{M}_{3+4}$ where $a$ and $b$ can be some
arbitrary function of the external momenta $p$, $k$, $p^{\prime }$ and $%
k^{\prime }$. In general, this would lead to some finite surface terms for
the difference of the linearly divergent integrals in the $i\mathcal{M}%
_{1+2} $ and $i\mathcal{M}_{3+4}$ as they actually are (see some discussion
below). However, in the dimensional regularization scheme taken here such
surface terms automatically vanish. So, the total contribution of \ four
diagrams (1+2+3+4) eventually comes to

\begin{eqnarray}
i\mathcal{M}_{1-4} &=&\frac{e^{3}n^{2}}{M}{\large [}\bar{u}(k^{\prime
})\gamma _{\lambda }u(p^{\prime })\cdot \bar{u}(k)\gamma _{\mu }u(p){\large ]%
}D^{\lambda \rho }(p-k)\cdot  \notag \\
&&\cdot \int \frac{d^{4}q}{(2\pi )^{4}}\left( \frac{D_{\rho }^{\mu }(q-k)}{%
(q-p)\cdot n}+\frac{D_{\rho }^{\mu }(q-p)}{(q-k)\cdot n}\right)  \label{1-4}
\end{eqnarray}
where we explicitly indicated the integration with respect to the internal
4-momentum $q$ \ (and used the symmetry of the propagator $D_{\rho }^{\mu
}(k-q)=D_{\rho }^{\mu }(q-k)$). One is allowed then to change internal
momentum $q$ to $-q$ in the second term in the integral in (\ref{1-4}) and
rewrite it as
\begin{equation}
I_{\rho }^{\mu }(p,k)=\int \frac{d^{4}q}{(2\pi )^{4}}\left( \frac{D_{\rho
}^{\mu }{\large (}q-k{\large )}}{{\large (}q-p{\large )}\cdot n}-\frac{%
D_{\rho }^{\mu }{\large (}q+(p+k)-k{\large )}}{{\large (}q+(p+k)-k{\large )}%
\cdot n}\right)  \label{dif}
\end{equation}
so that one has (again) the difference of two similar integrals in the $%
I_{\rho }^{\mu }(p,k)$ which only differs by the integration variables $q$
and $q+(p+k)$, respectively, that in the framework of the dimensional
regularization leads to their total cancellation.

Therefore, we have shown that the total matrix element $i\mathcal{M}_{tot}$
for the \ electron LIV scattering taken in the lowest $e^{3}/M$ order
including contribution of all eight diagrams (1-8) and given, as was said in
the above, by an extension
\begin{equation}
i\mathcal{M}_{tot}=i\mathcal{M}_{1-4}+(p\Leftrightarrow p^{\prime },\text{ \
}k\Leftrightarrow k^{\prime },\text{ \ }m\Leftrightarrow m^{\prime })
\end{equation}
does finally vanish in the dimensional regularization scheme.

\subsection{Integration: the surface term problem}

In conclusion, it seems interesting to discuss this LIV cancellation
mechanism from the surface term point of view in more detail, particularly,
for the integral $I_{\rho }^{\mu }(p,k)$ (\ref{dif}). What the physical LIV\
effect might be expected if the corresponding surface term were survived?
Note that, in the contrast to the above case with the $i\mathcal{M}_{3+4}$,
the shift in the integration variable in the $I_{\rho }^{\mu }(p,k)$ is
completely determined (as $(p+k)$) since both of integrals in it follow from
the same diagram (2) in the Fig.2. These integrals would, of course, exactly
cancel each other\ by the proper shift of the variables,\ if they were
finite. However, they, as one can readily see, are linearly divergent and,
generally speaking, such a shift of the integration variables is not
allowed. Instead, one should calculate the surface term in the $I_{\rho
}^{\mu }(p,k)$, as one usually does it when calculating the triangle anomaly
diagrams. And exactly, as in the anomaly case, this surface term appears
finite.

So, using the Gauss theorem one comes to
\begin{eqnarray}
I_{\rho }^{\mu }(p,k) &=&-\frac{(p+k)_{\tau }}{(2\pi )^{4}}\int d^{4}q\frac{%
\partial }{\partial q^{\tau }}\left( \frac{D_{\rho }^{\mu }{\large (}q-k%
{\large )}}{{\large (}q-p{\large )}\cdot n}\right) = \\
&=&-\frac{(p+k)_{\tau }}{(2\pi )^{4}}(2i\pi ^{2})\lim_{q\rightarrow \infty
}q^{\tau }q^{2}\left( \frac{-i{\large (}g_{\rho }^{\mu }-\frac{n^{\mu }%
{\large (}q-k{\large )}_{\rho }+n_{\rho }{\large (}q-k{\large )}^{\mu }}{%
n\cdot (q-k)}+\frac{n^{2}{\large (}q-k{\large )}^{\mu }{\large (}q-k{\large )%
}_{\rho }}{[n\cdot (q-k)]^{2}}{\large )}}{{\large (}q-p{\large )}^{2}{\large %
(}q-p{\large )}\cdot n}\right)  \notag
\end{eqnarray}
Now neglecting everywhere all the terms of the order $O(p/q,k/q)$ and using
for the limiting values of $q$ the evident equalities:
\begin{equation}
\frac{q^{\mu }q_{\rho }}{q^{2}}=\frac{1}{4}g_{\rho }^{\mu },\text{ \ }\frac{%
q^{\mu }q_{\rho }}{(n\cdot q)^{2}}=g_{\rho }^{\mu }\frac{1}{n^{2}},\text{ \
\ }\frac{n^{\mu }q_{\rho }+n_{\rho }q^{\mu }}{n\cdot q}=\frac{2n^{\mu
}n_{\rho }}{n^{2}},\text{ \ }\frac{q^{\tau }}{n\cdot q}=\frac{n^{\tau }}{%
n^{2}}
\end{equation}
one is eventually led to the finite value of the integral
\begin{equation}
I_{\rho }^{\mu }(p,k)=-\frac{(p+k)\cdot n}{4\pi ^{2}n^{2}}\left( g_{\rho
}^{\mu }-\frac{n^{\mu }n_{\rho }}{n^{2}}\right)
\end{equation}
as was expected.

The total matrix element $i\mathcal{M}_{tot}$ is now readily followed using
the orthogonality of the propagator $n_{\rho }D^{\lambda \rho }(p-k)=0$ so
that the last term in the $I_{\rho }^{\mu }(p,k)$ is not relevant. Taking
also that only the first term in the $D^{\lambda \rho }(p-k)$ contributes
when it is sandwiched between conserved fermion currents in the $i\mathcal{M}%
_{tot}$, and properly replacing momenta ($p\Leftrightarrow p^{\prime },$ \ $%
k\Leftrightarrow k^{\prime }$) to include all the contributions, one finally
comes to
\begin{equation}
i\mathcal{M}_{tot}=i\frac{e^{3}}{2\pi ^{2}}\frac{(p+p^{\prime })\cdot n}{M}%
{\large [}\bar{u}(k^{\prime })\gamma _{\mu }u(p^{\prime })\frac{1}{(p-k)^{2}}%
\bar{u}(k)\gamma ^{\mu }u(p)]  \label{tot}
\end{equation}
where we have also used the total 4-momentum conservation (giving $%
p+k+p^{\prime }+k\ ^{\prime }=2(p+p^{\prime })$) when collecting $i\mathcal{M%
}_{1-4}$ and $i\mathcal{M}_{5-8}$ $\ $in the $i\mathcal{M}_{tot}$.

Having the integrated LIV\ matrix element $i\mathcal{M}_{tot}$ (\ref{tot})
for the $ee^{\prime }$-scattering one can apply it (when properly modifying)
to any particular case, such as the $ee$ and $e\mu $ scatterings, $%
e^{+}e^{-} $-annihilation, $e^{+}e^{-}$ $\rightarrow \mu ^{+}\mu ^{-}$ \
conversion and so on, thus observing in the practical physics all the
peculiarities related with the Lorentz symmetry breaking. The common feature
for all these processes seems to be the direct dependence on some particular
component of the total momentum $(p+p^{\prime })\cdot n$ which for the case $%
M^{2}>0$ (fixed in the temporal choice of the vector $n^{\mu }=(1,0,0,0)$)
means the dependence on the total energy of $ee^{\prime }$-scattering.
Particularly, for the LIV correction to the Coulomb potential stemming from
the matrix elements $i\mathcal{M}_{tot}$ (\ref{tot}) in the non-relativistic
limit
\begin{eqnarray}
p &=&(m,\vec{p}),\text{ \ \ }k=(m,\vec{k}) \\
p\prime &=&(m\prime ,\vec{p}^{\prime }),\text{ \ \ }k=(m\prime ,\vec{k}%
\prime )  \notag \\
\bar{u}(k)\gamma _{\mu }u(p) &=&[\bar{u}(m)\gamma _{0}u(m),0,0,0]  \notag
\end{eqnarray}
(taken to the $O(\vec{p}^{2},\vec{k}^{2},\vec{p}^{\prime 2},\vec{k}^{\prime
2})$ accuracy) one has
\begin{equation}
i\mathcal{M}_{tot}=-i\frac{e^{3}}{2\pi ^{2}}\frac{(m+m^{\prime })}{M}{\large %
[}\bar{u}(m\prime )\gamma _{0}u(m^{\prime })\frac{1}{|\vec{p}-\vec{k}|^{2}}%
\bar{u}(m)\gamma _{0}u(m)]
\end{equation}
Collecting it with a standard QED matrix element (given by an ordinary
one-photon exchange) one is finally led to the total Coulomb potential for
the $ee\prime $ interaction
\begin{equation}
V_{el}(r)=\frac{e^{2}}{4\pi }\frac{1}{r}+\frac{e^{3}}{(2\pi )^{3}}\frac{1}{r}%
(\frac{m+m^{\prime }}{M})
\end{equation}
We have received in the certain sense somewhat exciting LIV extension of the
standard Coulomb potential. The extra term in it feels, as one can see, the
masses of the scattering particles and, thus, has some gravitational
character. At the same time its sign (repulsive or attractive) is determined
by the electromagnetism. Very remarkably, for the same sign charges just the
`anti-gravity' holds.

However, such a would-be finite physical LIV result depend, as one can see,
on some special condition for the virtual photon and fermion four-momentum
running in the loop diagrams in the Fig.2. Such a condition was taken in the
above so as to have the total reduction of the diagram (3) and (4) with the
proper parts from the diagram (2), while remaining some of its well defined
parts (\ref{1-4}). This corresponds just to the zero shifts ($a=b=0$) in the
integration variables in the integrals in the matrix element $i\mathcal{M}%
_{3+4}$ (\ref{3+4}) with respect to the integration variable in the
integrals in the $i\mathcal{M}_{1+2}$ (\ref{1+2}). Another condition could
give in principle another effect since the linearly (and higher) divergent
integrals are generally gauge and regularization scheme dependent. Point is,
however, that there happens to appear a freedom in the case considered to
choose the four-momentum running\ in the loops in a gauge invariant way not
to have the LIV at all. This is just what suggests the dimensional
regularization scheme.

\subsection{The other one-loop LIV contributions}

Following to the same way of argumentation we have also considered the other
leading one-loop LIV contributions. In the same order $e^{3}/M$ we have
calculated such contributions to the photon-photon and photon-fermion
scattering as well. Further, finding no the LIV \ \ \ corrections to photon
and fermion propagators we have checked then the proper one-loop
contributions to the fermion-fermion and photon-fermion scattering in the
next LIV order $1/M^{2}$. All these effects appears in fact vanishing due to
the same cancellation mechanism which we observed above for the tree and
loop LIV contributions: cancellation between diagrams containing the
3-photon vertex (where one or two of photons interacts with fermion in an
usual QED way) and those containing the contact 2-photon-fermion vertex.
Actually, their matrix elements (when they do not vanish by themselves as,
say, it takes place for the corrections to the photon and fermion
propagators) amount to the differences between pairs of the similar
integrals whose integration variables happen to be shifted relative to each
other by some constants (being in general arbitrary functions of the
external four-momenta of the particles involved) that in the framework of
the dimensional regularization leads to their total cancellation.

\section{Conclusion}

Some concluding remarks are in order:

1/ To the lowest LIV\ order (in $1/M$) all the tree diagrams caused by the
non-linear constraint $A_{\mu }^{2}=M^{2}$ in photon-photon, photon-fermion,
and fermion-fermion scatterings are exactly cancelled. In addition to the
Nambu's conclusion, we have shown that it has place for the both type of the
LIV, time-like ($M^{2}>0$) or space-like ($M^{2}<0$);

2/ It seems very likely that such tendency remains in the higher-order tree
diagrams as well, since there works the special mechanism of cancellation
between diagrams containing the 3-photon vertex (where one or two of photons
interacts with fermion in an usual QED way) and those containing the contact
2-photon-fermion vertex;

3/ For the one-loop diagrams this cancellation mechanism is also effective.
We have explicitly demonstrated it calculating the one-loop contributions to
the fermion-fermion scattering in the order of $e^{3}/M$ \ and also briefly
discussed the proper contributions to the photon-photon, photon-fermion and\
fermion-fermion scattering up to the higher LIV order $1/M^{2}$. All these
radiative effects appears in fact vanishing in the framework of the
dimensional regularization scheme taken;

4/ Together with all that the most important conclusion is that for the pure
QED like theories the standard potential-induced spontaneous symmetry
breaking (leading to the nonlinear field constraint $A_{\mu }^{2}=M^{2}$ or
to its more familiar linearized form $A_{\mu }=a_{\mu }+n_{\mu }M$) is in
fact superficial in the Lorentz symmetry case even in the case when quantum
corrections in terms of the one-loop contributions are included into
consideration. This happens to correspond only to fixing the gauge of the
vector potential in a special manner provided that its kinetic term is taken
in the standard gauge invariant $F_{\mu \nu }F^{\mu \nu }$ form. \

In this connection a critical question may appear in which extent our basic
conclusion related with the vanishing of all the one-loop LIV contributions
is the dimensional regularization scheme dependent, which was used in the
paper. We have shown in Sec.4.3 in somewhat provocative way how important
LIV modification of the Coulomb law could follow if some special condition
for the virtual photon and fermion four-momentum running in the loop
diagrams in the Fig.2. were taken. The point is, however, such a condition
would require another regularization scheme which could violate gauge
invariance in the theory and that scheme, therefore, could not be
implemented at higher loops to keep the theory unitary. The certain
advantage of the dimensional regularization scheme over the other ones is
that this scheme, preserving by itself the gauge invariance in the theory,
gives the direct and undistorted answer to the basic question we are
interested here - is the nonlinear field constraint $A_{\mu }^{2}=M^{2}$
only the gauge choice leading no to the observable Lorentz violating effects
in the otherwise gauge and relativistically invariant QED? And this answer
is positive at least in the one-loop approximation.

Nonetheless, whether it is possible to have the physical Lorentz violation
in some minimal extension of the conventional QED model. One way, proposed
Nambu\cite{nambu}, is, in fact, to add a term of type $\beta (\partial _{\mu
}A^{\mu })^{2}$ in the QED Lagrangian (\ref{lagr1}) which would lead to the
actual LIV in our nonlinear QED with three massless Goldstonic modes
appeared, two transversal and one longitudinal. This way, however, leads to
the uncontrollably large Lorentz violation since there is no reason to
consider the constant $\beta $ in the above bilinear term to be acceptably
small.

Another way was suggested recently in the paper\cite{cfmn}. It was shown
that starting with a general massive vector field theory one naturally comes
to the Lagrangian (\ref{lagr1}) with the nonlinear field constraint (\ref
{constr}) if the pure spin-1 value for the vector field $A_{\mu }(x)$ is
required. In essence, this model differs from the model considered in Sec.2
only in one substantial respect - the vector potential $A_{\mu }(x)$ appears
automatically satisfying to the transversality condition $\partial _{\mu
}A^{\mu }=0$. However, it happens enough for the physical LIV to occur.
Actually, as now one can see, due to the transversality condition the above
LIV cancellation mechanism is no longer working: the 3-photon (and generally
odd-number-photon) vertex disappears from theory so that the contact
2-photon-fermion (and even-number-photon-fermion in general) vertex diagram
contributions being proportional to the powers of the $1/M$ \ are left
uncompensated. This seems to be the only possible way how one could reach
the small and controllable physical breaking of Lorentz invariance\ in the
pure QED taken in the flat Minkowskian space-time. Unfortunately, such a
theory having now two field constraints appears in fact too restrictive in a
sense that the standard Coulomb law for the time-like LIV ($M^{2}>0$)
becomes problematic and should independently be introduced as a generic
four-fermion interaction. So, generally the LIV inspired QED with its
strictly conserved fermion current comes to the total conversion of the LIV
into gauge degrees of freedom of the massless photon thus coinciding with a
conventional QED. However, the situation is drastically changed when the
internal symmetry, together with the Lorentz invariance, is spontaneously
broken, as it takes place in Standard Model and Grand Unified Theories - the
LIV becomes physically observable\cite{cm}.

\section*{Acknowledgments}

We would like to thank Colin Froggatt, Rabi Mohaptra and Holger Nielsen for
useful discussions and comments which largely stimulated the present work.
Financial support from GRDF grant No. 3305 is gratefully acknowledged by
J.L.C.

\end{document}